\documentstyle[12pt]{article}
\begin{document}
\begin{titlepage}
\pagestyle{empty}
\topmargin-.5in
\baselineskip=14pt
\rightline{UMN-TH-1309/94}
\rightline{hep-ph/yymmddd}
\rightline{August 1994}
\vskip .2in
\baselineskip=18pt
\begin{center}
{\large{\bf  Light  Stops in the MSSM:
Implications for Photino Dark Matter and Top Quark Decay
}}
\end{center}

\vskip .1in
\begin{center}
Keith A. Olive and Serge Rudaz

{\it School of Physics and Astronomy, University of Minnesota}

{\it Minneapolis, MN 55455, USA}

\vskip .1in

\end{center}
\vskip .5in
\centerline{ {\bf Abstract} }
\baselineskip=18pt
We consider the viability of the minimal supersymmetric
standard model with a light ($m_{\tilde t_1} < $ 45 GeV) stop.
In order for its relic abundance to be cosmologically significant,
the photino as dark matter must be quite close in mass to the stop,
$(m_{\tilde t_1} - m_{\tilde \gamma}) \simeq  3 - 7 $ GeV. However, as
we show, the photino despite its low mass is virtually undetectable
by either direct or indirect means.
We also discuss the implications of these masses on the top quark
branching ratios.
\noindent
\end{titlepage}
%\newpage
\baselineskip=18pt

\def\la{~\mbox{\raisebox{-.6ex}{$\stackrel{<}{\sim}$}}~}
\def\ga{~\mbox{\raisebox{-.6ex}{$\stackrel{>}{\sim}$}}~}
\def\tm{$\tilde m$}
\def\phot{{\tilde{\gamma}}}
\def\t1{{{\tilde{t}}_1}}
\def\beq{\begin{equation}}
\def\eeq{\end{equation}}
\topmargin0.0in
\setcounter{footnote}{0}
Despite its name, the minimal supersymmetric standard model (MSSM)
contains a very large number of unknown mass parameters. Given some
theoretical assumptions, and the available accelerator constraints
on some of these masses, cosmology becomes  a useful  tool
in further constraining this parameter space
when the lightest supersymmetric particle (LSP) is stable
\cite {ehnos}. As the LSP is a potential dark matter candidate,
regions in parameter space can be excluded when the relic density of the
LSP is excessive. Though there are large portions of the total parameter
space in which the LSP in fact comprises much of the dark matter
necessary to obtain closure density (or the flat rotation curves of
spiral galaxies) \cite{mos,dn}, direct or indirect detection
of the LSP is most favorable when it is relatively light and
is a mix of higgsinos and gauginos. Light particles are preferable
since a fixed mass density provides a larger flux of
LSPs at low masses, while lower mass sfermions (to keep the cosmological
density down) make for larger elastic scattering cross-sections
(see. eg. \cite{pss}).

There is however a curious but surprisingly natural possibility that
all of the sfermions are in fact very heavy except for the lighter
of the two stop quarks \cite{er}. In this case, as has been
recently pointed out \cite{fmyy}, the LSP could very well be a light
photino (with $m_{\tilde \gamma} \sim 20$ GeV) with an acceptable
cosmological abundance. In what follows below, we determine the
extent to which the photino must be degenerate with the light
stop and examine the implications for the detection of such a light
LSP.  We find that despite its low mass, and the low mass of the
stop, this LSP is remarkably invisible to direct detection
and even to most methods of indirect detection.  Even the
potentially most favorable
possibility, which in this case seems to be the cold annihilation to
photon pairs in the galactic halo gives a signal well
below background.  We will also discuss the implications
of this scenario on the total width of the top quark relative
to the favored decay channel $t \rightarrow b W$. Light stops have also
 been considered recently in other contexts as well \cite{morestop}.

To easily see why one might expect a light stop, one needs
only to examine the
 general form of the sfermion mass matrix \cite{er}
\beq
\pmatrix{ \tilde f_L^* & \tilde f_R^* } ~
\pmatrix{ M_L^2 & m^2 \cr \noalign{\medskip} m^2 & M_R^2 \cr }~
\pmatrix{ \tilde f_L \cr \noalign{\medskip} \tilde f_R }\;,
\eeq
where $m^2=m _f(A_f + \mu \cot \beta)$ for weak isospin +1/2
fermions and $m_f (A_f + \mu \tan \beta)$ for weak isospin
$-1/2$. $A_f$ is the soft supersymmetry breaking trilinear mass term,
and $\mu$ is the Higgs mixing parameter; see \cite{gh} for
an overview of couplings in the MSSM. This mass matrix is easily
diagonalized
by writing the diagonal sfermion eigenstates as
\begin{eqnarray}
\tilde f_1 &=& \tilde f_L \cos\theta_f + \tilde f_R \sin\theta_f\;,\nonumber \\
\noalign{\medskip}
\tilde f_2 &=& - \tilde f_L \sin \theta_f + \tilde f_R \cos \theta_f\;.
\end{eqnarray}
With these conventions we have the  mass eigenvalues
\beq
m^2_{ 1,2 } =  { 1 \over 2 }  \left[ ( M_R^2 + M_L^2 ) \mp \sqrt {
( M_R^2 - M_L^2 )^2 +
4 m^4 } \right]\;.
\eeq
Note that in the special case $M_L = M_R = M$, we have
$\theta_f= $ sign[$-m^2$]$(\pi/4)$ and $ m^2_{ 1,2 } = M^2 \mp |m^2|$.

Now let us suppose that all of the supersymmetry mass scales ($M_R, M_L$,
and $A_f$) are large ($\ga 300$ GeV) except for the gaugino masses
which remain relatively light. For the SU(2) gaugino mass $M_2
\la 50$ GeV and $\mu \la 300$ GeV, the LSP is predominantly a
photino (see eg. \cite{os34}). Note however that for large $\tan
\beta (\ga 3)$ or for $\mu > 0$ (unless $\tan \beta \la 1.2$)
much of this parameter space is excluded due to the experimental
absence
of light charginos.  Thus for large $A_f$, it is quite plausible
that $m^2$ is comparable to $M_R$ and $M_L$ for tops since
$m_t \approx 174$ GeV \cite{top}.
As we will see, this region of parameter space is allowed, and it
 provides us with a relatively light cold dark matter
candidate (a photino) which
surprisingly would be very difficult to detect.

     The light scalar top ${\tilde t}_1$ pair coupling to the
 $Z^o$  is proportional to   ${1 \over 2} \cos^2 \theta_t -
{2 \over 3} \sin^2 \theta_W$
and vanishes for $\theta_t \simeq 0.98$ rad.
 For this value of $\theta_t$, direct
searches for ${\tilde t}_1$ at   $e^+e^-$ machines rely
on photon mediated production.  A
published report \cite{venus} by the VENUS Collaboration at
TRISTAN ($\sqrt{s} = 58$ GeV)
excludes ${\tilde t}_1$ in the mass range between 7.6 and 28.0 GeV,
for nearly all
values of $m_\phot$ , except those very close to the kinematical
limit for the
dominant decay ${\tilde t}_1 \rightarrow c \phot$ (in fact,
 the bound does not
depend on the precise identification of the lightest neutralino).
There are
no published reports of similar searches at LEP,
although preliminary results
from the DELPHI Collaboration have appeared in conference
 proceedings \cite{delphi}.
The results allow a light scalar top below $M_Z/2$ provided that
$0.9 \la \theta_t \la 1.1$, and that ${\tilde t}_1$ and $\phot$
    have masses that are not too different,
which is the case we will consider here.

In order to avoid an excessive cosmological mass
density, a photino with mass $m_{\phot} < 35$ GeV would
normally require sfermion masses $m_{\tilde{f}} \la 140$
GeV.  For larger sfermion masses, as we are considering here,
the photino annihilation cross sections will be too small
to maintain a relic density $\Omega_\phot h^2 \la 1$.
The fact that we have a light stop is of no help because
annihilation into a $t, \bar t$ pair is kinematically
forbidden. However, because photinos remain in thermal equilibrium
with the stops (if they are close in mass) through the
process $\phot + c
\leftrightarrow {\tilde t}_ 1$ (for light stops in the mass range considered
 in this paper, the radiatively induced process ${\tilde t}_ 1
\rightarrow \phot + c$ is the dominant stop decay mode \cite{tcpho}),
the annihilation of stops
can ensure a sufficiently low relic density of photinos \cite{fmyy}.
In what follows below, we will refine this calculation and determine the
allowable photino mass as a function of the light stop mass.
We will also discuss the potential (or lack thereof) for detecting
this light photino.  Finally, we will discuss the implications
of these parameters on the top quark branching ratios.

Although the photino annihilation cross section is too small
to reduce the relic abundance of photinos, the decays and
inverse decays of ${\tilde t}_ 1 \leftrightarrow c + \phot$
keep photinos in equilibrium with stops whose annihilation cross
section is then sufficient for reducing the number density of both
\cite{gs}. From the rate ${\tilde t}_ 1 \rightarrow c + \phot$
\cite{tcpho}
\beq
\Gamma({\tilde t}_ 1 \rightarrow c + \phot) =
(0.3 - 3) \times 10^{-10} m_{{\tilde t}_1} \left(
1 - {{m_\phot}^2 \over  {m_{{\tilde t}_1}}^2 } \right)^2
\eeq
it is easy to see that when compared to the expansion rate
of the Universe,
\beq
H = {8 \pi \over 3} N^{1/2} T^2/M_P
\eeq
where $N$ is the number of relativistic degrees of freedom at temperature
$T$, photinos and stops will always be in equilibrium at $T \la 10^5$
GeV, unless stops and photinos are {\em extremely} degenerate.
Therefore to determine the relic density of photinos one needs
only the annihilation cross section of stops into gluons \cite{fmyy}
(the co-annihilation of photinos and stops  to charm plus glue
is sufficiently suppressed that it can be ignored in this context)
\beq
\sigma v = {14 \pi \alpha_s^2 \over 27 {m_{{\tilde t}_1}}^2}
\eeq
The Boltzmann equation for this system is most simply given in terms
of the sum of the number densities $n = n_\phot + n_\t1$ \cite{swo,gs}
\beq
{dn \over dt} = - 3 H n - \langle \sigma_{\rm eff} v \rangle
\left( n^2 - {n_{\rm eq}}^2 \right)
\label{be}
\eeq
where the effective cross-section is defined to be
\beq
\langle \sigma_{\rm eff} v \rangle =  2 \sigma v r^2 \qquad
r = {n_\t1 \over n} \simeq {3 \over 2} \left( {m_\t1 \over
m_\phot} \right)^{3/2} e^{-\Delta / x}
\eeq
with $\Delta = (m_\t1 - m_\phot)/m_\phot$ and $x \equiv T/m_\phot$.

It is relatively straightforward to integrate the Boltzmann
equation (\ref{be}). The result is simply
\beq
n = \left( {4 \pi^3 N \over 45} \right)^{1/2} {T^3 \over m_\phot
M_P} \left( \int_{x_f}^\infty \langle \sigma_{\rm eff} v \rangle
dx \right)^{-1}
\eeq
where $x_f$ corresponds to the temperature at freeze-out
determined by the condition
\beq
{.0046 \over \sqrt{N}} \left( {{m_\t1} M_P \over
{m_\phot}^2 } \right) x^{-1/2} e^{-(2\Delta + 1)/x}
\simeq 1
\eeq
Integrating the cross-section, gives
\beq
\int_{x_f}^\infty \langle \sigma_{\rm eff} v \rangle
dx \simeq 0.21 \left( \Delta m_\t1 \over {m_\phot}^3 \right)
\Gamma(-1,2\Delta /x_f) \simeq {0.05 \over \Delta} \left({{m_\t1} \over
{m_\phot}^3} \right) e^{-2 \Delta  / x_f} x_f^2
\left( 1 - {x_f \over \Delta } \right)
\eeq
so that
\beq
\Omega_\phot h^2 = {m_\phot n \over \rho_c} = 1.1 \times 10^{-10}
\left( { ({m_\phot}/{\rm GeV})^3 \over \Delta \Gamma(-1,2 \Delta/x_f)}
\right) \left({T_\phot \over T_\gamma}\right)^3 \sqrt{N}
\eeq
The result for the relic photino density is shown in figure 1 as a
function of the photino mass for different values of the stop mass.
(Note that for the curve labeled 20, corresponding to the choice of the
stop mass, only the portion of the curve above $\sim 17$ Gev
is allowed by VENUS results \cite{venus}.)
These results are qualitatively similar (and close quantitatively)
to those in ref. \cite{fmyy}.  It is clear that for a wide range
of photino masses between 17 and 33 GeV, the relic density is significant
so long as the stop mass is close (within 3 to 7 GeV) to that
of the photino mass.

Let us now turn to the potential for detecting this light
photino.  Detection of the LSP can be divided among
direct methods (typically involving cryogenic laboratory detectors)
\cite{pss,ef}
and indirect methods which make use of the annihilation of LSP's
trapped in the sun or the earth \cite{sos} or
annihilations in the halo of the galaxy \cite{gamma}-\cite{gamma3}.
The direct detection methods as well as the indirect methods
looking for the annihilation products of trapped LSP's, depend crucially
on the elastic scattering cross-section of LSP's on matter.
For pure (or nearly pure) photinos, as is the case we are considering
here, the elastic cross sections are suppressed
by ${m_{\tilde f}}^4$ where ${m_{\tilde f}}$ is the mass of the heavy
and potentially very heavy squarks. Since the top quark content
of the proton is negligible, the light stop makes virtually
no contribution to the elastic scattering cross section.
Typically, for elastic scattering mediated by sfermions, the scattering
cross section leads to detection rates which are in fact marginal
when $m_{\tilde f} \simeq 100 GeV$.  For the sfermion masses we consider here,
these rates would be lowered by two orders of magnitude or more.

For heavier LSP's, $m_\chi > 35$ GeV, (LSP's with larger values of
$M_2$ and similar values of $\mu$ compared to the photinos considered here),
the LSP, is no longer a pure photino, but rather becomes more
of a generic mix between a gaugino and higgsino \cite{os34}.
In this case, if there is in addition a sufficiently light
Higgs boson, the elastic scattering may be enhanced \cite{fos}.
However, for $m_\phot \la 35$ GeV, we have the curious situation
where we are allowed a light photino (if the stop is nearly as light)
which at the same time, if the all the other sfermions are very heavy,
is essentially completely transparent to matter.

 An indirect method of detection of relic photinos as a dominant component
of the Galactic Halo that is also of interest is to search for a narrow line
in the cosmic gamma ray spectrum at  $E_\gamma  = m_\phot$, produced by the
annihilation reaction $\phot \phot \rightarrow \gamma \gamma$
in the halo \cite{gamma1}-\cite{gamma3}.  Four types of
Feynman diagrams contribute to this process (together with their crossed
partners) as shown in  Fig.
2, where the internal lines correspond to
either scalars (dashed) or fermions (full). We consider the contributions
of all fermions but the
top quark, for which $m_f \ll m_{\tilde f}$, separately.
 For these fermions, the
dominant contribution to the annihilation amplitude is found to be of order
$(eQ_f)^4m_\phot^2/m_{\tilde f}^2$: the contribution of a heavy top quark with
$m_t \simeq m_\t1$ is negligible.
The observability of the gamma ray line flux under
these conditions has been considered in detail in \cite{gamma3}:
as this flux scales like ${m_{\tilde f}}^{-4}$,
the signal dwindles into insignificance if values of ${m_{\tilde f}}$ of
several hundred GeV are considered.

     The contribution of the top quark to this process must however be
reconsidered in the light stop scenario, $m_t \gg m_\t1$.  Indeed, if all
other scalar superpartners are taken to be much heavier, one could hope that
top intermediate states could still lead to an observable signal.  This,
unfortunately, is not what happens.  In the light stop scenario, the leading
contribution from top intermediate states will come from diagrams 2(c) and
2(d),
which together form a gauge invariant subset.
Their contribution is reliably calculated in the local limit
$m_\t1 \ll m_t$ by first deriving the effective lagrangian
 ${\cal L}_{\rm eff}$ for $\phot \phot \rightarrow \t1 \t1^*$
 and obtaining the $\phot \phot \rightarrow \gamma \gamma$
 matrix element
\beq
{\cal M}(\phot \phot \rightarrow \gamma \gamma) = \langle \gamma
\gamma | {\cal L}_{\rm eff} | \phot \phot \rangle
\eeq
The effective Lagrangian for $\phot \phot \rightarrow \t1 \t1^*$
 has the leading terms:
\beq
{\cal L}_{\rm eff} = {4 e^2 \over 9 m_t} \left[ {m_\phot \over m_t}
- \sin 2 \theta_t \right] \bar \lambda (x) \lambda(x) \t1^\dagger(x)
\t1(x) + {2 e^2 \over 9 m_t^2} \cos 2 \theta_t
\bar \lambda (x) \gamma_\mu \gamma_5 \lambda(x) {\tilde J}^\mu(x)
\eeq
where $\lambda(x)$ is the (Majorana) photino field and
${\tilde J}^\mu(x)$ the vector current
of the scalar tops,
\beq
{\tilde J}^\mu(x) =  ~i~\t1^\dagger(x)
{\stackrel{\leftrightarrow}{\partial^\mu}} \t1(x)
\eeq
Now one can quickly see what results for the
$\phot \phot \rightarrow \gamma \gamma$ process:  the
matrix element is written as,
\beq
\langle \gamma
\gamma | {\cal L}_{\rm eff} | \phot \phot \rangle =
{4 e^2 \over 9 m_t} \left[ {m_\phot \over m_t}
- \sin 2 \theta_t \right] \langle \gamma \gamma |\t1^\dagger \t1
 | 0 \rangle \langle 0 | \bar \lambda \lambda | \phot \phot \rangle
+ {2 e^2 \over 9 m_t^2} \cos 2 \theta_t \langle \gamma \gamma |
{\tilde J}^\mu(x) |0 \rangle \langle 0 |
 \bar \lambda \gamma_\mu \gamma_5 \lambda |
\phot \phot \rangle
\label{supp1}
\eeq
The second term vanishes identically by C-invariance,
$\langle \gamma \gamma |
{\tilde J}^\mu(x) |0 \rangle \equiv 0$, since the current
${\tilde J}^\mu$ is C-odd, while the vacuum and two-
photon states are C-even.
In the first term, the matrix element $\langle
 \gamma \gamma |\t1^\dagger \t1
 | 0 \rangle$ does not
vanish, and is readily evaluated: one finds,
\beq
\langle \gamma(k_1 \gamma(k_2) | \t1^\dagger \t1
 | 0 \rangle = {2 e^2 \over 3 \pi^2 m_\t1^2} ( \eta^{\mu\nu}
k_1 \cdot k_2 - k_1^\nu k_2^\mu) \epsilon_\mu^*(k_1)
\epsilon_\nu^*(k_2) I(m_\phot/m_\t1)
\eeq
where
\beq
I(\xi) = \int_0^1 dy \int_0^1 dx {x^3 y (1-y) \over 1 - 4 \xi^2 x^2 y (1-y)}
\eeq
This integral is fairly easily expressible in closed form: we shall not bother
here, because now
 it is the second factor of the first term that suppresses
this matrix element enormously.  Indeed, for non-relativistic
$(\beta = v/c \simeq 10^{-3})$
relic photinos in the halo, the matrix element of the scalar
 density is found to be suppressed,
\beq
\langle 0 | \bar \lambda \lambda | \phot \phot \rangle \simeq
{\rm O}(m_\phot \beta)
\label{supp2}
\eeq
As a result, the amplitude due to stop exchange behaves parametrically as
$\beta m_\phot^3/m_\t1^2 m_t$, to be compared with the light fermion
contribution
discussed earlier which goes like  $m_\phot^2/m_{\tilde f}^2$.
 It is clear without
any further ado (numerical factors are easily taken into account) that given
the range of $m_\t1$ considered here and the necessary constraint on
$m_\phot/m_\t1$, the top contribution is always substantially
less than that of the light
fermions, even for the largest conceivable values of $m_{\tilde f}$,
in which case
the line signal is all but unobservable anyway, as mentioned before.
     Given that the leading contribution from diagrams 2(c) and 2(d) with
top intermediate states is p-wave suppressed (Eqs. \ref{supp1},\ref{supp2}),
 we need to discuss the
contributions of diagrams 2(a) and 2(b) as well. Here, the dominant
contribution
to the amplitude comes from loop momenta of order $m_t$ in the box graphs, so
one cannot take the local limit before doing the loop integrals.
Furthermore, in the limit $m_\t1^2 \rightarrow 0$ (i.e. $m_\t1^2
\ll m_t^2$), the diagrams 2(a) and 2(b) are convergent in the infrared,
so unlike the case of diagrams 2(c) and 2(d), there will be no inverse
powers of $m_\t1^2$. Now the
leading behavior can be obtained by the informed use of dimensional analysis.
Under the most favorable circumstances, the parametric dependence of the
matrix element from these diagrams would be $m_\phot^3/m_t^3$:
whether or not there
is also a p-wave suppression factor can only be decided by explicit
calculation.  If $m_{\tilde f} \simeq 2 m_t$,
 say, this could be of the same order of
magnitude as the $m_\phot^2/m_{\tilde f}^2$
 of a given light fermion: while this would be
larger than the contribution from the other pair of diagrams, the gamma ray
line flux would still be too low to be observable.

     Thus, we come to the conclusion that in the light stop scenario, with
multi-hundred GeV scalar superpartners, the
$\phot \phot \rightarrow \gamma \gamma$ process in the
Halo could not lead to an observable cosmic gamma ray line signature.

The light stop scenario has many implications in the MSSM, for example
as regards the $\rho$
parameter, $K^o - {\bar K^o}$ mixing, the rare decay $b \rightarrow s \gamma$,
proton decay through dimension-5 operators in MSGUT, and generally
the light sparticle spectrum.  A review of these issues in \cite{fmyy}
concludes that this scenario is perfectly allowed by known
phenomenology.  Here, in view of the recent announcement by the CDF
collaboration \cite{top} of evidence for the top quark we wish to consider
only the implications of the light stop scenario on top decay branching
ratios.
 In addition to the canonical decay channel
\beq
\Gamma(t\rightarrow bW) = {G_F m_t^3 \over 8 \sqrt{2} \pi } \left(
1 - {M_W^2 \over m_t^2} \right)^2 \left( 1 + {2 M_W^2 \over m_t^2} \right)
= 1.53 {\rm GeV}
\eeq
there is the possibility for decays into photinos and stops with a rate
given by \cite{er}
\beq
\Gamma(t\rightarrow \t1 \phot) = {\alpha \over 9 m_t^3} (m_t^2 + m_\phot^2 -
m_\t1^2 - 2 m_t m_\phot \sin 2\theta_t ) \lambda^{1/2}(m_t^2, m_\phot^2,
m_\t1^2) \theta(m_t-m_\phot-m_\t1)
\label{decay}
\eeq
For stops with $m_\t1 \la 32 GeV$, (assuming the GUT relation between
the photino and gluino masses) there is also the possibility that
tops decay to stops and gluinos with a rate given by eq. (\ref{decay})
with the substitution $4\alpha/9 \rightarrow 4\alpha_s/3$ and
$m_\phot \rightarrow m_{\tilde g}$.
(We implicitly assume that the decay $t \rightarrow b H^+$, a possibility
in the MSSM, is absent due to kinematics, i.e. we assume that the $H^+$ is
heavy.)
 The branching ratio for the
$t\rightarrow bW$ decay is shown in figure 3 as the area between the two
curves. In computing these curves,
we have assumed the GUT relation between the gaugino masses
so that $m_{\tilde g} = 5.76 m_\phot$ and we have taken the value of $\theta_t$
to be that for which the coupling of the light stop
to the $Z^o$ vanishes. The upper curve corresponds to the branching ratio
when $(m_\t1 - m_\phot) = 2 - 3$ GeV (the minimum allowed from figure
1 for this range of stop masses) and the lower curve corresponds to
the branching ratio
when $(m_\t1 - m_\phot) = 3 - 7$ GeV (the maximum allowed from figure
1 for this range of stop masses).
As one can see, for $m_\t1 \ga 30$ GeV, the only additional channel
is the decay to stops and photinos which makes a 5 \% contribution
to the total width. At lower values of $m_\t1$ the contribution from
$t \rightarrow \t1 \tilde g$ becomes significant.
The reduced $t \rightarrow b W$ branching ratio would then worsen
the discrepancy \cite{top} between the CDF top quark signal and the
expected QCD rate for $p\bar p \rightarrow t \bar t X$ production.
Thus, it would appear that such considerations would disfavor the low range
of mass values $m_\t1 \la 30$ GeV.
However, it should be borne in mind that gluinos in the corresponding mass
range,
up to $\sim 145$ GeV will be pair produced in $p \bar p$
collisions, with a sizable branching ratio for the decay
${\tilde g} \rightarrow \t1 {\bar t} ({\rm virtual}) \rightarrow
(c \phot) (\bar b W)$, providing a new source of events
with $b$ jets and $W$'s which should be taken into account in data analysis.

\vskip 0.8truecm
\noindent {\bf Acknowledgments}
\vskip 0.4truecm
We would like to thank J. Cline, K. Kainulainen, M. Srednicki,
and M. Voloshin for
helpful conversations.
This work was supported in part by the U.S.
 Department of Energy under contract
DE-FG02-94ER-40823.

\newpage
\noindent{\bf{Figure Captions}}

\vskip.3truein

\begin{itemize}

\item[]
\begin{enumerate}
\item[]
\begin{enumerate}

\item[{\bf Figure 1:}]  The relic density of photinos as a
function of the photino mass. The different curves are
labeled by the stop quark mass in GeV.

\item[{\bf Figure 2:}] Typical box diagrams contributing to the
transition amplitude for  the process $ \phot \phot \rightarrow
\gamma \gamma$. The corresponding diagrams with photon lines
interchanged are not shown.

\item[{\bf Figure 3:}] The branching ratio for $t \rightarrow b W$
as a function of the stop mass.  The two curves encompass the allowed
range of photino masses which give a cosmologically allowed
and interesting relic density from figure 1.

\end{enumerate}
\end{enumerate}
\end{itemize}

\end{document}